\documentclass[runningheads]{svmult}

\usepackage{makeidx}   % allows index generation
\usepackage{graphicx}  % standard LaTeX graphics tool
\usepackage{subeqnar}  % subnumbers individual equations
\usepackage{multicol}  % used for the two-column index
\usepackage{physprbb}  % modified textarea for proceedings,
\makeindex             % used for the subject index
\begin{document}
\title*{Convective Instability of Magnetized Ferrofluids:
\protect\newline
Influence of Magnetophoresis and Soret Effect}
\toctitle{Convective Instability of Magnetized Ferrofluids:
\protect\newline
Influence of Magnetophoresis and Soret Effect}
\titlerunning{Instability of Magnetized Ferrofluids}
\author{Mark I. Shliomis}
\authorrunning{Mark I. Shliomis}
\institute{Department of Mechanical Engineering,
       Ben-Gurion University of the Negev,\\
       P.O.B. 653, Beer-Sheva 84105, Israel}

\maketitle              % typesets the title of the contribution

\begin{abstract}
\index{Convective instability} Convective instability in a
ferrofluid layer heated from below or from above in the presence
of a uniform vertical \index{magnetic field} magnetic field is
investigated theoretically. Convection is caused by a magnetic
mechanism based on the temperature and concentration dependence of
magnetization. An imposed temperature gradient establishes (by the
{\em Soret effect}) a concentration gradient of magnetic particles
of which the ferrofluid is composed. Both these gradients cause a
spatial variation in \index{magnetization} magnetization, which
induces a gradient of magnetic field intensity within the fluid
layer. The field gradient induces in its turn an additional
redistribution of \index{magnetic grains} magnetic grains due to
{\em magnetophoresis}. Resulting self-consistent \index{magnetic
force} magnetic force tries to mix the fluid. A linear stability
analysis predicts {\em oscillatory instability} in a certain
region of the magnetic field strength and the fluid parameters.
The instability owes magneto- and thermophoresis its origin: were
the particle diffusion not operative, then only stationary
instability would occur. A discovery of predicted convective
oscillations is expected in ferrofluid layers about $1\,{\rm mm}$
thick, where the buoyancy mechanism is negligible and the
characteristic diffusion time is not too long.
\end{abstract}

\section{Introduction}

\index{Magnetic colloids} Magnetic colloids (so-called {\em
ferrofluids} [1,2]) represent very interesting object for studies
of \index{convection} convection. Even in the absence of the
gravity, mechanical equilibrium of a nonisothermal ferrofluid in a
magnetic field $H$ is in general impossible. At the basis of the
mechanism of {\em thermomagnetic} convection [1-5] lies the
temperature dependence of the magnetization $M$: under otherwise
equal conditions, the colder elements of the fluid are more
strongly magnetized, and therefore they are subject to a larger
magnetic force in the direction of $\nabla H$. The gradients of
the magnetic field intensity here play the same role as does the
gravitational field $\bf g$ in the mechanism of ordinary
thermogravitational convection, whilst the temperature dependence
of the magnetization, $M(T)$, plays the role of $T$-dependence of
the fluid density $\rho$.

Interestingly, \index{thermomagnetic convection} thermomagnetic
convection can arise even in a {\em uniform} applied magnetic
field [4]. Let a ferrofluid confined between two horizontal planes
be in a uniform vertical external field ${\bf H}_{\rm e}=(0, 0,
H_{\rm e})$ at the temperature $T(z)$. Then the dependence $M(T)$
leads to the result that the field {\em inside} the fluid, ${\bf
H}=(0, 0, H)$, appears to be {\em nonuniform}. Indeed, from the
Maxwell equation ${\rm div}\,{\bf B}=0$ and the definition of
magnetic induction ${\bf B=H}+4\pi{\bf M}$ it follows
$$\frac{{\rm d}H}{{\rm d}z}=-\,4\pi\,\frac{{\rm d}M}{{\rm d}z}\,,\eqno(1.1)$$
that yields
$$\frac{{\rm d}H}{{\rm d}z}=-\frac{4\pi}{\hat{\mu}}\,\Biggl(\frac{\partial M}
{\partial T}\Biggr)_{\!\!\rm H}\frac{{\rm d}T}{{\rm
d}z}\,,\eqno(1.2)$$ where $\hat{\mu}=1+4\pi(\partial M/\partial
H)$ is the differential \index{magnetic permeability} magnetic
permeability. Equation (1.1) shows that the induced gradient of
magnetic field is directed always {\em opposite} to the gradient
of magnetization. It results in unstable magnetization
stratification: the magnetic force ${\bf F}=M\nabla H$ tries to
mix ferrofluid. According to eq. (1.2), this force is proportional
to the imposed temperature gradient. Therefore, when both the
magnetic and buoyancy mechanisms are operative, the first of two
predominates over the second one at the onset of convection in a
{\em thin enough} fluid layers, where the critical temperature
gradient is sufficiently large. One can show [1] that the
condition
$$\Bigg|\Biggl(\frac{\partial M}{\partial T}\Biggr)_{\!\!\rm H}\frac{{\rm d}H}{{\rm d}z}
\Bigg|\gg\Bigg|\Biggl(\frac{\partial \rho}{\partial T}\Biggr)_{\!\!\rm p}g\Bigg|\,,$$
which permits neglect the buoyancy mechanism of convection, is reduced to
$$d^4\ll 4\pi Ra_{\rm c}\eta\kappa\Biggl[\frac{(\partial M/\partial T)_{\rm H}}
{g(\partial\rho/\partial T)_{\rm p}}\Biggr]^2, \eqno(1.3)$$ where
$d$ is the layer thickness, $Ra_{\rm c}\sim 10^3$ the critical
\index{Rayleigh number} Rayleigh number ($Ra_{\rm c} =1708$ in the
case of rigid boundaries and $Ra_{\rm c} =27\pi^4/4\approx 657.5$
for free boundaries), $\eta$ and $\kappa$ the fluid viscosity and
thermodiffusivity. For commonly used magnetite ferrofluids on the
base of water or kerosene the \index{pyromagnetic coefficient}
pyromagnetic coefficient $-(\partial M/\partial T)_{\rm H}$ in the
field  $H\sim 200-500\,\rm Oe$ is about $0.5\phi\,\,\rm G/K$ where
$\phi$ is the volume fraction of magnetic grains. Then for
$\phi\sim 0.1,\,\,\,\eta\sim 10^{-2}{\rm Ps},\,\,\,\kappa\sim
10^{-3}\, \rm cm^2/s$, and $-(\partial\rho/\partial T)_{\rm p}\sim
5\times 10^{-4}\,\rm g\,cm^{-3}\,K^{-1}$,  the inequality (1.3)
holds true down to $d=1\,\rm mm$. The same conclusion has been
done by Finlayson [4] who first studied theoretically convective
instability of a ferrofluid layer heated  from below in the
presence of a uniform vertical magnetic field.

It is important to notice that in [4] and posterior theoretical
works [5-8] ferrofluids were treated as pure (i.e.,
single-component) magnetized fluids. Such a model assumes a fixed
uniform concentration of magnetic grains, $\phi=\rm const$.
Ignoring in that way {\em diffusion effects}, authors of [4-8]
came to conclusion that only {\em stationary} convective
instability is possible in ferrofluids. As it pointed out already
in the abstract of Ref. [4], "oscillatory instability cannot
occur". But a real ferrofluid should be treated as a {\em binary
mixture} with allowance for concentration dependence of its
magnetization. An externally imposed temperature gradient induces
in a ferrofluid layer a concentration gradient owing to the {\em
Soret effect}. Both these gradients cause a spatial variation in
magnetization $M(\phi, T, H)$ that leads in turn to appearance of
a magnetic field gradient within the fluid layer. Thus there arise
magnetic forces, which try to mix the ferrofluid and, in addition,
they induce a certain redistribution of magnetic grains due to
\index{magnetophoresis} {\em magnetophoresis}. As it will be shown
below, just the magnetophoresis together with the thermophoresis
(the \index{Soret effect} Soret effect) give rise to {\em
oscillatory instability} in a certain region of the magnetic field
intensity and the fluid parameters. Appropriate experimental
conditions to observe predicted instability take place in a thin
ferrofluid layer, when the ordinary (buoyancy) mechanism of
convection is negligible and the characteristic diffusion time,
$\tau_{\mathrm D}=d^2/\pi^2D$, required to settle a steady state,
is not too long (here $D$ is the diffusion coefficient of magnetic
particles in the basic liquid). Recent measurements [9] yielded
$D=2.7\times 10^{-7}\,\rm cm^2/s$ in water and cyclohexane. (This
magnitude of $D$ agrees well with a theoretical estimate by the
Einstein's formula for the diffusion coefficient, $D=k_{\mathrm
B}T/6\pi\eta a$, where $2a\simeq 10\,\rm nm$ is the diameter of a
magnetic grain). According to above formula for $\tau_{\mathrm
D}$, the concentration equilibrium in a ferrofluid layer of the
thickness $d=1\,\rm mm$ is reached after one hour.

In the next Section we obtain the description of the
magnetophoresis and Soret effect in a ferrofluid assuming that the
fluid behaves with respect to an external magnetic field like a
Langevin's paramagnet [1,10]. In Secs. 3 and 4 we derive the
complete set of modified linear equations for perturbations and
settle the boundary conditions for them. The eigenvalue problem is
solved in Sec. 5 for the B\'{e}nard configuration. On the base of
the solution, oscillatory instability is predicted to occur for a
wide range of magnitudes of magnetic field, Soret coefficient, and
other fluid parameters.

Owing to the small magnitude of $D$ (see above), there are possible two scenarios of the
beginnings of convection. Depending on the imposed heating rate, the ferrofluid may behave
either as a binary mixture or like a pure fluid. Both the possibilities are discussed in Sec. 6.

\section{Magnetophoresis and Soret Effect}

In a nonuniform magnetic field there acts on each particle a force ${\bf f=(m\nabla)H}$.
Orientation of the particle magnetic moment $\bf m$ in the direction of $\bf H$ is impeded by
thermal motion. As a result, the mean value of $\bf m$ appears to be equal to
$$\langle{\bf m}\rangle = m {\cal L}\left(mH\over{k_{\mathrm B}T}\right){\bf e}\;,\;\;\;
{\bf e}=\frac{{\bf H}}{H} \;,\;\;\;{\cal L}(\xi)=\coth \xi-\xi^{-1} \;.\eqno(2.1)$$
With allowance for the identity ${\bf H\times{\rm rot} H}=\frac{1}{2}\nabla H^2-
\bf (H\nabla)H$ and the condition of absence of electric currents in nonconducting ferrofluids,
${\rm rot}\,\bf H=0$, formula (2.1) enable us to write the mean magnetophoretic force
$\langle{\bf f}\rangle=m\,{\cal L}({\xi})\nabla H.$ This force and the Stokes drag coefficient,
$6\pi\eta a$, for a sphere of the radius $a$ determine the regular component of the Brownian
velocity of the particle with respect to the liquid,
$${\bf u}=\frac{m}{6\pi\eta a}{\cal L}({\xi})\nabla H.\eqno(2.2)$$

The volume density of the diffusion flux of the matter in a ferrofluid may be written
$${\bf j}=\phi{\bf u}-D({\nabla \phi+S_{\mathrm T}\nabla T}),\eqno(2.3)$$
where $S_{\mathrm T}$ is the Soret coefficient. Substituting (2.2) in (2.3) and using the
Einstein's formula for the diffusion coefficient, we obtain
$${\bf j}=-D[\nabla \phi+S_{\mathrm T}\nabla T-(\phi/H)\xi{\cal L}({\xi})\nabla H].\eqno(2.4)$$
Thus the matter flux is provided by the "diffusiophoresis" ($\propto \nabla \phi$; this term
was proposed by Derjaguin et al. [11]), thermophoresis ($\propto \nabla T$), and magnetophoresis
($\propto \nabla H$). If, however, confined horizontal planes are impervious and an applied
magnetic field is uniform, then gradients of $\phi$ and $H$ can be caused only by an imposed
temperature gradient. Therefore in the steady state (${\bf j}=0$) the values of $\nabla \phi$
and $\nabla H$ are expressed via $\nabla T$. As it is seen from (2.4), the equilibrium equation
$j_{\mathrm z}=0$ reads
$$\frac{{\rm d}\phi}{{\rm d}z}=-S_{\rm T}\frac{{\rm d} T}{{\rm d}z}+\phi\xi{\cal L}\,
\frac{{\rm d}\ln H}{{\rm d}z}\,.\eqno(2.5)$$
Substituting in (1.1) $M=\phi M_{\rm s}{\cal L}(mH/k_{\rm B}T)$ where $M_{\rm s}$ is the
saturation magnetization of the particle material, we find
$$\hat{\mu}\frac{{\rm d}H}{{\rm d}z}=-4\pi\Biggl(\frac{{\rm d}\ln\phi}{{\rm d}z}-
\frac{\xi{\cal L'}}{\cal L}\frac{{\rm d}\ln T}{{\rm d}z}\Biggr).\eqno(2.6)$$
Eliminating now the concentration gradient from (2.5) and (2.6), we arrive at
$$\frac{{\rm d} H}{{\rm d}z}=\frac{4\pi M}{\sigma T}\Biggl(\Psi+\frac{\xi{\cal L'}}
{\cal L}\Biggr)\frac{{\rm d} T}{{\rm d}z}\,,\eqno(2.7)$$ where
$\sigma=\hat{\mu}+12\pi\chi_0{\cal L}^2(\xi)=1+12\pi\chi_0({\cal
L'+L}^2),\,\,\, \chi_0=\phi M_{\rm s}m/3k_{\rm B}T$ the initial
Langevin \index{magnetic susceptibility} magnetic susceptibility,
and $\Psi=(T/\phi)S_{\rm T}$ plays the role of the {\em separation
ratio} in the mechanism of thermomagnetic convection. Substituting
${\rm d} H/{\rm d}z$ from (2.7) into (2.5) we obtain, finally,
$$\frac{{\rm d}\phi}{{\rm d}z}=-\frac{\hat{\mu}\phi}{\sigma T}\Biggl[\Psi-\frac{(\hat{\mu}-1)}
{\hat{\mu}}\,\xi{\cal L}\Biggr]\frac{{\rm d} T}{{\rm d}z}\,.\eqno(2.8)$$
One can rewrite the latter in the form
$$\frac{{\rm d}\phi}{{\rm d}z}=-\frac{\phi}{T}\Psi_{\rm H}\frac{{\rm d} T}{{\rm d}z}\,,
\eqno(2.9)$$ introducing the field-dependent \index{separation
ratio} separation ratio
$$\Psi_{\rm H}=\frac{\hat{\mu}}{\sigma}\Biggl[\Psi-\frac{(\hat{\mu}-1)}{\hat{\mu}}\,
\xi{\cal L}\Biggr].\eqno(2.10)$$
Note that $\Psi_{\rm H}$ differs from zero even if $\Psi=0$, i.e., in the absence of the Soret
effect. The {\em non-Soret} $\phi-T$ coupling originates from the magnetophoresis, which is
manifested as a negative separation ratio:
$$\Psi_{\rm H}|_{\Psi=0}=-(\hat{\mu}-1)\,\xi{\cal L}/\sigma<0.\eqno(2.11)$$
Actually, the induced gradient of magnetic field (2.7) is directed along the temperature
gradient, thus the hotter fluid (and hence the stronger magnetic field) -- the higher
concentration of magnetic grains. By the same manner it behaves a nonmagnetic binary mixture
with a negative separation ratio. The function defined by Eq. (2.11) is proportional to
$\xi^2$ at $\xi\ll 1$ and inversely proportional to $\xi$ at $\xi\gg 1$. For a ferrofluid with
the initial magnetic permeability $\mu_0=1+4\pi\chi_0=3$ this function has a distinct minimum,
$\Psi_{\rm H}|_{\Psi=0}=-0.303$, at $\xi=2.30$.

\section{Equations for Stability Analysis}

The general expression for the volume density of magnetic force acting a nonconducting and
incompressible ferrofluid may be written [12]
$${\bf F}=\nabla\Biggl\{\frac{H^2}{2}\Biggl[\rho\Bigg(\frac{\partial \chi}
{\partial\rho}\Biggr)_{\!\mathrm {T,\,H}}-\chi\Biggr]\Biggr\}+M\nabla H,\eqno(3.1)$$
where $\chi=M/H$ stands for the magnetic susceptibility. The first term in (3.1) is known as
the {\em magnetostrictive force} and the second one called the {\em Kelvin force}. Under the
substitution $\bf F$ in the equation of fluid motion, the magnetostrictive force which
originates from a potential is included in the pressure gradient, $\nabla p$, so this force is
equilibrated {\em automatically}. Thus it remains only the Kelvin force, which in equilibrium
should be also balanced at each point by the pressure gradient: $\nabla p=M\nabla H$. On
applying the operation $\rm rot$ to this equation, we obtain a necessary condition for
equilibrium,
$$\Biggl(\frac{\partial M}{\partial T}\nabla T+\frac{\partial M}{\partial \phi}\nabla
\phi\Biggr)\times\nabla H=0.\eqno(3.2)$$
In the case under consideration, all three gradients contained in (3.2) are parallel to each
other (they have only $z$-components), so the condition (3.2) is satisfied. Thus, the
equilibrium is possible, and we set about a study of its stability.

Small perturbations of a standing mode will be characterized by velocity $\bf v$, pressure $p$,
temperature $\theta$, concentration $\varphi$, and magnetic field $\bf h$. Then the perturbation
${\bf F'}$ of the magnetic force density ${\bf F}=M\nabla H$ is expressed via $\theta,\;
\varphi$ and $\bf h$:
$${\bf F'}=-h_{\mathrm z}\nabla M+\Biggl(\frac{\partial M}{\partial T}\theta+\frac{\partial M}
{\partial \phi}\varphi+\frac{\partial M}{\partial H}h_{\mathrm z}\Biggr)\nabla H.\eqno(3.3)$$
[We have omitted here the term $\nabla(Mh_{\mathrm z})$. Later on it is assumed to be included
in the gradient of the pressure perturbation, $\nabla p$, see Eq. (3.6)]. In the equilibrium
$\nabla M$ and $\nabla H$ are directed along the $z$ axis; so, using (1.1) we obtain from (3.3)
the only component of the vector ${\bf F'}$:
$$F'_{\mathrm z}=\Biggl(\frac{\hat{\mu}}{4\pi}h_{\mathrm z}+\frac{\partial M}{\partial T}
\theta+ \frac{\partial M}{\partial \phi}\varphi\Biggr)\frac{{\rm d}H}{{\rm d}z}\,.\eqno(3.4)$$
Substituting in the formula $M=\phi M_{\mathrm s}{\cal L}(\xi),\;\xi=mH/k_{\rm B}T$, and
${\rm d}H/{\rm d}z$ from (2.7), we get
$$F'_{\mathrm z}=\frac{M}{\sigma T}\Biggl(\Psi+\frac{\xi{\cal L'}}{\cal L}\Biggr)
\Biggl[\hat{\mu}h_{\mathrm z} +4\pi M\Biggl(\frac{\varphi}{\phi}-\frac{\xi{\cal L'}}
{\cal L}\frac{\theta}{T}\Biggr)\Biggr]  \frac{{\rm d}T}{{\rm d}z}\,.\eqno(3.5)$$

Thus, the linear equation of fluid motion takes the form
$$\rho\frac{\partial{\bf v}}{\partial t}=-\nabla p+\eta\nabla^2{\bf v}+\frac{M\Delta T}
{\sigma Td}\Biggl(\Psi+\frac{\xi{\cal L'}}{\cal L}\Biggr)\Biggl[4\pi M\Biggl(\frac{\varphi}
{\phi}-\frac{\xi{\cal L'}}{\cal L}\frac{\theta}{T}\Biggr)+\hat{\mu}\frac{\partial \Phi}
{\partial z}
\Biggr]{\bf e}.\eqno(3.6)$$
We have substituted in (3.6) the magnitude of the stationary temperature gradient,
${\rm d}T/{\rm d}z=\Delta T/d$, where $\Delta T$ is the temperature difference between confined
surfaces of the layer. In (3.6) $\bf e$ is the unit vector along the $z$ axis, and $\Phi$ the
potential of magnetic field perturbations, ${\bf h}=\nabla\Phi$. To derive an equation for
$\Phi$ one should substitute the perturbation of magnetic induction,
$${\bf b=h}+4\pi\frac{M}{H}({\bf h}-h_{\mathrm z}{\bf e})+4\pi\Biggl(\frac{\partial M}
{\partial T}\theta+\frac{\partial M}{\partial \phi}\varphi+\frac{\partial M}{\partial H}
h_{\mathrm z}\Biggr){\bf e},$$
in the Maxwell equation $\rm div\,{\bf b}=0$. It yields
$$\mu\nabla^2\Phi-(\mu-\hat{\mu})\frac{\partial^2\Phi}{\partial z^2}=-4\pi M\Biggl(\frac{1}
{\phi}\frac{\partial \varphi}{\partial z}-\frac{\xi{\cal L'}}{T{\cal L}}\frac{\partial \theta}
{\partial z}\Biggr),\eqno(3.7)$$
where $\mu=1+4\pi(M/H)=1+12\pi\chi_0{\cal L}(\xi)/\xi$.

Equation (2.4) under the replacement $\phi\rightarrow \phi+\varphi,\,\,\,\,T\rightarrow T+
\theta$, and $H\rightarrow H+\partial\Phi/\partial z$ determines the matter flux perturbation.
So, making allowance for Eq. (2.9), we obtain the diffusion equation
$$\frac{\partial \varphi}{\partial t}=D\nabla^2 \Biggl(\varphi+\frac{\phi}{T}\Psi\theta-
\xi{\cal L} \frac{\phi}{H}\frac{\partial \Phi}{\partial z}\Biggr)+\frac{\phi\Delta T}{Td}
\Psi_{\rm H}\,({\bf ve}). \eqno(3.8)$$
Equations (3.6)-(3.8) together with the heat conduction equation,
$$\frac{\partial \theta}{\partial t}=\kappa\nabla^2\theta-(\Delta T/d)({\bf ve}),\eqno(3.9)$$
and the condition of incompressibility, $\rm div\,{\bf v}=0$, form the complete set of
equations which determine the onset of convection in a magnetized ferrofluid.

Pass over to dimensionless variables by choosing a unit of length $d$, time
$d^2/\kappa$, velocity $\kappa/d$, temperature $\Delta T$, concentration $(\Delta T/T)\phi$,
and field potential $4\pi M (\Delta T/T)d$. Then the above equations take the form
$$Pr^{-1}\frac{\partial{\bf v}}{\partial t}=-\nabla p+\nabla^2{\bf v}+Rm\,G\Biggl[\Omega-
\Biggl(\Psi+\frac{\xi{\cal L'}}{\cal L}\Biggr) \theta+\sigma{\rm D}\Phi\Biggr]{\bf e},
\eqno(3.10\rm a)$$
$$\frac{\partial \theta}{\partial t}=\nabla^2\theta-({\bf ve}),\;\;\;\;\;
{\rm div}\,{\bf v}=0,\eqno(3.10\rm b)$$
$$\frac{\partial \Omega}{\partial t}=Le\nabla^2\Omega+\Psi\frac{\partial \theta}{\partial t}-
(\sigma-\hat{\mu})\,{\rm D}\frac{\partial \Phi}{\partial t}+\Psi_{\mathrm H}\,({\bf ve}),
\eqno(3.10\rm c)$$
$$\mu\nabla^2\Phi+(\sigma-\mu)\,{\rm D}^2\Phi=-{\rm D}\Biggl[\Omega-\Biggl(\Psi+
\frac{\xi{\cal L'}}{\cal L}\Biggr)\theta\Biggr]\,,
\;\;\;\;\;\;\;\;\;\; \nabla^2\Phi_{\rm e}=0\;.\eqno(3.10\rm d,
e)$$ Here $Pr=\eta/\rho\kappa$ is the \index{Prandtl number}
Prandtl number, $Le=D/\kappa$ the \index{Lewis number} Lewis
number, ${\rm D}\equiv{\rm d/\,d}z$,
$$Rm=\frac{[\phi M_{\rm s}\,\Delta T\,d]^2}{\eta\kappa\,T^2},\;\;\;\;\;\;\;G=\frac{4\pi
{\cal L}^2}{\sigma}\Biggl(\Psi+\frac{\xi{\cal L'}}{\cal
L}\Biggr),\eqno(3.11{\rm a, b})$$ where $Rm$ we shall call the
\index{magnetic Rayleigh number} magnetic Rayleigh number,
$\Phi_{\rm e}$ is the potential of field perturbations outside the
fluid layer, and
$\Omega=\varphi+\Psi\theta-(\sigma-\hat{\mu})\,{\rm D}\Phi$ is
introduced as an independent variable instead of the concentration
perturbation $\varphi$. It is worth to note that $\nabla\Omega$
has the physical meaning of the matter flux perturbation.

Studying the onset of convection in the B$\acute{\rm e}$nard configuration, one should put all
perturbations be dependent on time and horizontal coordinates as $\exp[\lambda t+
\imath(k_{\mathrm x}x+k_{\mathrm y}y)]$. Then from Eqs. (3.10) it follows the system of
equations for $z$-dependent amplitudes of $z$-component of velocity $W=({\bf ve})$,
temperature $\theta$, matter flux potential $\Omega$, and magnetic field potentials $\Phi$ and
$\Phi_{\mathrm e}$:
$$({\rm D}^2-k^2)({\rm D}^2-k^2-\lambda/Pr)W-k^2Rm\,G\,[\Omega-
(\Psi+\xi{\cal L'/L})\theta+\sigma{\rm D}\Phi]=0\;,\eqno(3.12\rm a)$$
$$({\rm D}^2-k^2-\lambda)\theta-W=0\;,\eqno(3.12\rm b)$$
$$Le({\rm D}^2-k^2-\lambda/Le)\Omega+\lambda\Psi\theta-(\sigma-\hat{\mu}) \lambda{\rm D}\Phi+
\Psi_{\rm H}W=0\;,\eqno(3.12\rm c)$$
$$(\sigma{\rm D}^2-\mu k^2)\Phi=-{\rm D}[\Omega-(\Psi+\xi{\cal L'/L})\theta],
\;\;\;\;\;\;\;({\rm D}^2-k^2)\Phi_{\rm e}=0\;.\eqno(3.12\rm d,e)$$

\section{Boundary Conditions}

To completely specify solutions of Eqs. (3.12), one needs ten
\index{boundary conditions} boundary conditions at confined
surfaces with the addition of two conditions on $\Phi_{\mathrm e}$
far from the layer.

For the case of rigid boundary surfaces whose heat conductivity is very high in comparison with
the enclosed ferrofluid, the boundary conditions on velocity and temperature read:
$$W={\rm D}W=\theta=0\;\;\;\;\;\;\;\;{\rm at}\;\;\;\;z=\pm 1/2\;.\eqno(4.1)$$
It is natural to assume such boundaries to be at the same time completely impervious. The latter
condition is expressed by setting the matter flux perturbation equal to zero,
$${\rm D}\Omega =0\;\;\;\;\;\;\;\;{\rm at}\;\;\;\;z=\pm 1/2\;.\eqno(4.2)$$
Boundary conditions on the magnetic potential $\Phi$ are pretty complicated by the fact that
$\Phi$ induces outside the layer a magnetic potential  $\Phi_{\rm e}$ governed by Eq.
(3.12\rm e). This equation have the solution $\Phi_{\rm e} \propto \exp(\pm kz)$ which must
decay far from the layer and be valid everywhere outside the layer including the layer
boundaries too. Hence we come to the relations
$${\rm D}\Phi_{\mathrm e}=-k\Phi_e\;\;\;\;{\rm at}\;\;\;\;z=1/2\,; \;\;\;\;\;\;\;\;\;\;\;\;
{\rm D}\Phi_{\mathrm e}=k\Phi_e\;\;\;\;{\rm at} \;\;\;\;z=-1/2\;.\eqno(4.3)$$
Boundary conditions of continuity of the tangential components of magnetic field and the normal
component of magnetic induction on the layer boundaries have the form
$$\partial\Phi/\partial x=\partial\Phi_e/\partial x\,,\;\;\;\;\;\;\;\;
\partial\Phi/\partial y=\partial\Phi_e/\partial y\;,\eqno(4.4)$$
and
$${\rm D}\Phi+4\pi\Biggl(\frac{\partial M}{\partial \phi}\varphi +\frac{\partial M}{\partial T}
\theta+\frac{\partial M}{\partial H}{\rm D}\Phi\Biggr)={\rm D} \Phi_{\rm e}\;.\eqno(4.5)$$
The conditions (4.4) are satisfied if
$$\Phi=\Phi_{\rm e}\;\;\;\;\;\;{\rm at}\;\;\;\;\;\;z=\pm 1/2\;,\eqno(4.6)$$
whereas Eq. (4.5) on the substitution $M=M(\phi,T,\xi)$ takes the dimensionless form
$$\sigma{\rm D}\Phi+\Omega -\Biggl(\Psi+\frac{\xi{\cal L'}}{\cal L}\Biggr)\theta ={\rm D}
\Phi_{\rm e}\;\;\;\;\;\;{\rm at}\;\;\;\;\;\;z=\pm 1/2\;.\eqno(4.7)$$
Relations (4.3) and (4.6) permit to write down: ${\rm D}\Phi_{\rm e}=\mp k\Phi\;$ at $\;
z=\pm 1/2$, that gives us an opportunity to eliminate ${\rm D} \Phi_{\rm e}$ from (4.7). Taking
into account the boundary conditions on the temperature (4.1), we obtain, finally, the closed
boundary conditions on $\Phi$:
$$\sigma{\rm D}\Phi+k\Phi=-\Omega\;\;\;{\rm at}\;\;\;z=1/2\,;\;\;\;\;\;\;\;\; \;\;
\sigma{\rm D}\Phi-k\Phi=-\Omega\;\;\;{\rm at}\;\;\;z=-1/2\,.\eqno(4.8)$$
Thus, there is no necessity at all to find the magnetic potential outside the layer. The
relations (4.1), (4.2), and (4.8) represent the complete set of the rather realistic boundary
conditions for the set of equations (3.12${\rm a}$)--(3.12${\rm d}$).

Below, however, we consider the case of idealized boundary conditions
$$W={\rm D}^2W=\theta=\Omega={\rm D}\Phi=0\;\;\;\;\;\;{\rm at}\;\;\;\;\;\;z=\pm 1/2\;,
\eqno(4.9)$$
which would be true for free, perfectly heat conducting and pervious boundaries made of a
superconductor. In the latter case, magnetic field perturbations do not penetrate into the bulk
of constrained horizontal plates, so that ${\rm D}\Phi_{\mathrm e}=0$. Thus, as it follows
from (4.7) with allowance for the boundary conditions (4.9) on $\theta$ and $\Omega$, the
value ${\rm D}\Phi$ at $z=\pm 1/2$ appears to be zero too.

\section{Exact Solution for Free Boundaries}

In the case of boundary conditions (4.9) there exists an analytical solution of Eqs.
$(3.12{\rm a})-(3.12{\rm d})$. Satisfying the boundary conditions and the equations by means of
$$W\propto\theta\propto\Omega\propto\cos\,\pi z\,, \;\;\;\;\;\;\;\;\Phi\propto\sin\,\pi z\,,$$
we arrive at the characteristic equation for increments $\lambda$,
$$(\pi^2+k^2+\lambda)(\pi^2+k^2+\lambda /Pr)(\pi^2+k^2)[\lambda (\alpha \pi^2+k^2)+
Le(\beta\pi^2+k^2)(\pi^2+k^2)]$$
$$-k^4Rm\,G\,[\lambda (\xi{\cal L'/L})+(\pi^2+k^2+\lambda)\Psi_{\mathrm H}+ Le(\pi^2+k^2)
(\Psi+\xi{\cal L'/L})]=0\;,\eqno(5.1)$$
where $\alpha =\hat{\mu}/\mu$ and $\beta =\sigma /\mu$. For stationary convective instability
($\lambda =0$) the magnetic Rayleigh number is given by
$$Rm^{\rm st}=\frac{Le\,(\beta\pi^2+k^2)(\pi^2+k^2)^3}{k^4G\,[\Psi_{\rm H}+Le\,(\Psi+
\xi{\cal L'/L})]}\;.$$
As it mentioned above, magnetic colloids are characterized by extremely small Lewis numbers
(typically $Le\sim 10^{-4}$). Therefore, using the definitions (2.10) and $(3.11{\rm b})$, one
may rewrite the last equation in the form
$$Rm^{\rm st}=\frac{Le\,\sigma^2\,(\beta\pi^2+k^2)\,(\pi^2+k^2)^3} {4\pi\hat{\mu}k^4{\cal L}^2
\,(\Psi+\xi{\cal L'/L})[\Psi-(\hat{\mu}-1)\xi{\cal
L}/\hat{\mu}]}\;.\eqno(5.2)$$ The magnetic Rayleigh number
according to its definition $(3.12{\rm a})$ is always {\em
positive}. Therefore, as it seen from (5.2), \index{stationary
instability} stationary instability sets in only {\em outside} the
interval
$$-\frac{\xi{\cal L'}(\xi)}{{\cal L} (\xi)}<\Psi<\frac{\hat{\mu}-1}{\hat{\mu}}\,\xi{\cal L}
(\xi)\;, \eqno(5.3)$$ i.e., either for a large enough positive
separation ratio or for a sufficiently small negative $\Psi$. At
the same time, as it will be shown below, {\em inside} the
interval it can occur only {\em oscillatory} convective
instability. In particular, the \index{oscillatory instability}
oscillatory instability is only possible in the absence of the
Soret effect, i.e., when $\Psi =0$.

Omitting in (5.1) small terms with the coefficient $Le$, we obtain the cubic equation for
$\lambda$:
$$(\alpha\pi^2+k^2)(\pi^2+k^2)\lambda^3+(Pr+1)(\alpha\pi^2+k^2)(\pi^2+k^2)^2\lambda^2$$
$$+[(\alpha\pi^2+k^2)(\pi^2+k^2)^3-k^4Rm\,G\,(\Psi_{\mathrm H}+\xi{\cal L'/L})]Pr\lambda$$
$$= k^4(\pi^2+k^2)Pr\,Rm\,G\,\Psi_{\mathrm H}\;.\eqno(5.4)$$
Substituting in this equation $\lambda =\I \omega$ we arrive at two relations which determine
the magnetic Rayleigh number for oscillatory instability,
$$Rm^{\rm osc}=\frac{(Pr+1)\,\sigma^2\,(\alpha\pi^2+k^2)\,(\pi^2+k^2)^3}{4\pi\hat{\mu}Pr\,
k^4{\cal L}^2(\Psi+\xi{\cal L'/\cal L})[\Psi+(\xi{\cal L'/L})(1+\sigma/\hat{\mu}\,Pr)]}\;,
\eqno(5.5)$$
and the frequency of neutral convective oscillations,
$$\omega^2=-\frac{[\Psi-(\hat{\mu}-1)\xi{\cal L}/\hat{\mu}]\,(\pi^2+k^2)^2} {\Psi+
(\xi{\cal L'/L})(1+\sigma/\hat{\mu}\,Pr)}\;.\eqno(5.6)$$
Since $Rm^{\rm osc}$ and $\omega^2$ must be both positive, the oscillatory instability can
occur only if $\Psi$ lies within the interval (5.3). The corridor of oscillatory instability
for a ferrofluid with the initial magnetic permeability $\mu_0=3$ is plotted in Fig. 1 for a
wide enough range of the Langevin parameter $\xi$. On the upper border of the corridor, the
frequency of convective oscillations turns into zero, whereas on its lower border the critical
Rayleigh number turns into infinity. In the absence of magnetophoresis the corridor grows
narrow so that its upper  border passes along the line $\Psi =0$. In this case the oscillatory
instability would be possible  owing to the Soret effect only for $\Psi<0$. At last, if neither
magnetophoresis nor Soret effect are in operation (i.e., {\em no particle diffusion}), then the
cubic equation (5.4) for $\lambda$ is reduced to the square one,
$$\hat{\mu}\,(\alpha\pi^2+k^2)\,(\pi^2+k^2)\,[\lambda^2+(Pr+1)\,(\pi^2+k^2)\,
\lambda +Pr\,(\pi^2+k^2)^2]$$
$$=4\pi Pr\,k^4\,Rm\,(\xi{\cal L'})^2.\eqno(5.7)$$
This equation has no solutions of kind of $\lambda =\I \omega$. Thus, if there is no diffusion,
then there is no oscillatory instability as well. For stationary instability  Eq. (5.7) gives
$$Rm^{(0)}=\frac{\hat{\mu}\,(\alpha\pi^2+k^2)\,(\pi^2+k^2)^3}{4\pi
k^4 (\xi{\cal L'})^2}\;.\eqno(5.8)$$

\begin{figure}
\begin{center}
\includegraphics[width=.9\textwidth]{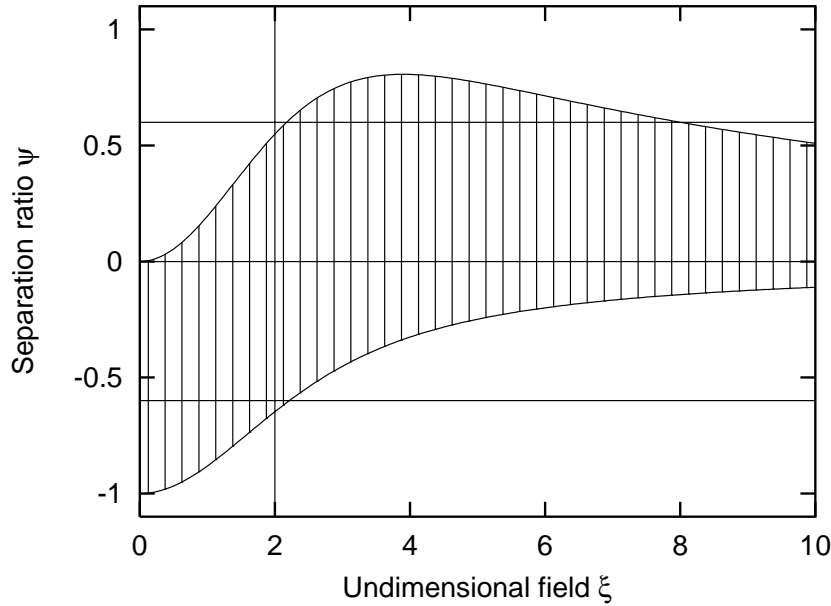}
\end{center}
\caption{The corridor of oscillatory instability ({\em dashed}) in $(\xi\,,\Psi)$ plane.
{\em Straight lines} correspond to $\Psi=0,\;\Psi=-0.6,\;\Psi=+0.6$ and $\xi=2$. Cross sections
of the surface $Rm_{\rm c}(\xi, \Psi)$ along these lines are presented below in Figs. 2-5
respectively.}
\label{f1}
\end{figure}

Below we shall consider $Rm$ as a function of $\xi$ and $\Psi$. Any pair of fixed magnitudes of
the parameters determines a neutral stability curve in the $(Rm\,,\;k)$ plane. The minima of
these curves, $Rm_{\rm c}$, and the critical circular frequency of convective oscillations,
$f=\omega_{\rm c}/2\pi$, are presented in Figs. 2-5. Figures 2-4 show $Rm_{\rm c}$ and $f$ as
functions of $\xi$ for three magnitudes of $\Psi$ marked in Fig. 1 by horizontal lines
$\Psi =0$, $\Psi=-0.6$, and $\Psi =+0.6$. Figure 5 represents a stability diagram in the plane
$(Rm_{\rm c},\;\Psi)$ for the magnitude $\xi =2$ marked in Fig. 1 by the vertical line.

\section{Discussion of Results and Conclusion}

At first, let us consider the case $\Psi =0$. Then Eqs. (5.5)-(5.6) become somewhat simpler:
$$Rm^{\rm osc}=\frac{(Pr+1)\,\sigma^2\,(\alpha\pi^2+k^2)\,(\pi^2+k^2)^3}{4\pi k^4\,
(\xi{\cal L'})^2\,(Pr\,\hat{\mu}+\sigma)}\;,$$
$$\omega^2=\frac{3Pr\,(\mu_0-1){\cal L}^2\,
(\pi^2+k^2)^2}{Pr\,\hat{\mu}+\sigma}\;.\eqno(6.1)$$ The function
$Rm^{\rm osc}(k)$ has a minimum $Rm_{\mathrm c}^{\rm osc}$ at the
\index{critical wave number} critical wave number $k_{\mathrm c}$
determined by the formula
$$k_{\rm c}^2=(\pi/2)^2\,(1-\alpha+\sqrt{1+14\alpha+\alpha^2})\;.\eqno(6.2)$$
Here $\alpha =\hat{\mu}/\mu\leq 1$ depends on the dimensionless magnetic field $\xi$ and the
initial magnetic permeability $\mu_0$. For commonly used ferrofluids (with $1<\mu_0< 5$)
$k_{\rm c}$ weakly depends on $\xi$. For instance, at $\mu_0=3$ the critical wave number lies
inside the interval $2.97<k_{\rm c}<3.14$ for any $\xi$, so that $k_{\rm c}$ variations do not
exceed $6\%$.

Substituting $k_{\rm c}^2$ from (6.2) into (6.1) we find the critical Rayleigh number and the
frequency of neutral oscillations. The values $\log Rm_{\rm c} ^{\rm osc}$ and
$f=\omega_{\rm c}/2\pi$ as functions of $\xi$ are presented in Fig. 2. In the same figure it
is shown the logarithm of $Rm_{\rm c}^{(0)}$ determined by Eq. (5.8) under the replacement
$k^2$ by $k_{\rm c}^2$ from (6.2).

\begin{figure}
\begin{center}
\includegraphics[width=.9\textwidth]{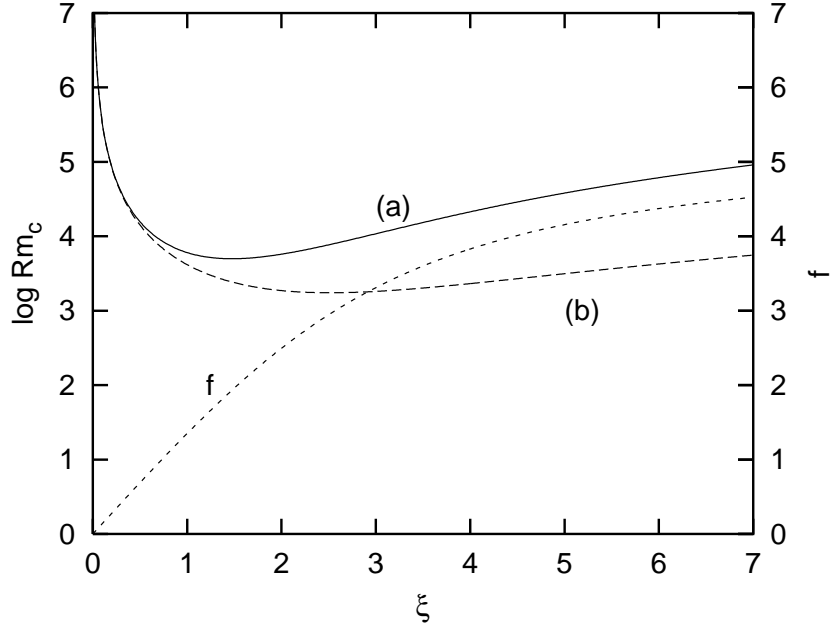}
\end{center}
\caption{Critical magnetic Rayleigh number $Rm_{\rm c}$ and circular frequency of neutral
oscillations $f$ versus magnetic field strength $\xi$. Curve $(a)$ -- oscillatory solution
$Rm_{\rm c}^{\rm osc}$ obtained with allowance for magnetophoresis but without Soret effect
($\Psi=0$), curve $(b)$ -- stationary solution $Rm_{\rm c}^{(0)}$ in the absence of the
particle diffusion (i.e., there are neither magnetophoresis nor the Soret effect).}
\label{f2}
\end{figure}

We dare say that both the critical Rayleigh numbers, $Rm_{\rm c}^{\rm osc}$ and
$Rm_{\rm c}^{(0)}$ have a direct physical meaning. Depending on conditions of a real experiment
it will set in either stationary convection above $Rm_{\rm c}^{(0)}$ or oscillatory convection
above $Rm_{\rm c}^{\rm osc}$. The point is that ferrofluids are characterized by two very
different time scales [13]. One of two is the above-mentioned long mass diffusion time
$\tau_{\rm D}= d^2/\pi^2D$: even in a thin layer ($\sim 1\,{\rm mm}$) the concentration
equilibrium is established for a time longer than {\em one hour}. Another one is a short
thermodiffusion time $\tau_{\rm T}=d^2/\pi^2\kappa$: in the same fluid layer the stationary
temperature gradient sets in merely in {\em one second}. The ratio $\tau_{\rm T}/\tau_{\rm D}$
is equal just to the Lewis number $Le=D/ \kappa\sim 10^{-4}$. The presence of such different
time scales makes possible two scenarios of the onset of convection. First, the temperature
difference between the layer boundaries can be increased from zero till a supercritical value
during a time $\tau\ll\tau_D$. Then diffusion processes have no time to evolve, so the magnetic
colloid behaves like a pure fluid. Indeed, magnetic grains do not move without diffusion
relative to the surrounding liquid matrix, they are "frozen" into it. Just that very case has
been studied in [3-8]. Stationary convection starts in this case at $Rm_{\mathrm c}^{(0)}$. The
latter value has a minimum at a certain $\xi\,$; for a water-based ferrofluid ($Pr=7$) with
$\mu_0=3$ this minimum, $\min[Rm_{\mathrm c}^{(0)}]=1742$, is reached at $\xi=2.52$ and
$k_{\rm c}=3.02$.

Second scenario can be realized in a thin enough fluid layer, where the long diffusion time,
$\tau_{\rm D}$, is {\em not too long}. Then it is possible to adjust so low heating rate, that
the temperature difference across the layer, $\Delta T$, will be formed for a time
$\tau>\tau_{\rm D}$. Under this condition the concentration gradient induced by the temperature
gradient due to magnetophoresis (the Soret effect is absent at $\Psi=0$) will be built up
undisturbed by convection. In this case, as it shown above, there arises oscillatory instability
at $Rm_{\rm c}^{\rm osc}$. For a ferrofluid under consideration the critical Rayleigh number
reaches the minimum, $\min [Rm_{\rm c}^{\rm osc}]=4993$, at $\xi=1.47$; corresponding wave
number and frequency of neutral oscillations are $k_{\rm c}=3.08$ and $\omega_{\rm c}=12.13$.
The frequency increases monotonously with $\xi$ approaching the saturation value
$\omega_{\rm c}=34.19$ at $\xi\rightarrow \infty $.

Let us estimate the temperature difference and other parameters required to cause oscillatory
convection by the thermomagnetic mechanism. Assuming the mean diameter of magnetite particles
($M_{\mathrm s}=480\,{\rm G}$) to be $2a=11\,{\rm nm}$, we obtain the particle magnetic moment
$m=(4\pi a^3/3)M_{\rm s}=3.345\times 10^{-16}\,{\rm erg/Oe}$, so the value
$\mu_0=1+4\pi\phi M_{\rm s}m/3k_{\rm B}T=3$ is reached at the room temperature at $\phi=12\%$,
and $\xi=mH/k_{\rm B}T=1$ at $H=122\,{\rm Oe}$. Thus the minimum in the curve of
$\log\,Rm_{\rm c}^{\rm osc}$ in Fig. 2 takes place at $H=180\,{\rm Oe}$ (to get such a field
within the ferrofluid layer one must apply externally $H_e=\mu_0H=540\,{\rm Oe}$). The minimum
Rayleigh number ($\approx 5000$) is reached at $(\Delta T)_c=(1.02/d)\,{\rm K/cm}$, thus the
critical temperature difference is about $10\,{\rm K}$ in the layer of $d=1{\rm mm}$. For this
layer the unit of time, $d^2/\kappa$, is equal to $6.9\,{\rm s}$, so the above-pointed
dimensionless frequency $\omega_{\rm c}=12.13$ of critical convective oscillations corresponds
to the period of $3.6\,{\rm s}$.

Maps of stability in the $(Rm_{\rm c},\,\xi)$--plane are presented in Figs. 3 and 4 for
$\Psi=-0.6$ and $\Psi=+0.6$, respectively. As it seen in Fig. 1, for both these magnitudes of
the separation ratio there are certain intervals of $\xi$ in which it can occur oscillatory
instability. Out of the intervals there arises stationary Soret convection which is
characterized by very small Rayleigh numbers (5.2) due to the smallness of the Lewis number
(for a water-based ferrofluid $Le=1.9\times 10^{-4}$). The critical Rayleigh number for
stationary instability, $Rm_{\rm c}^{\rm st}$, is obtained from (5.2) under the substitution
$k_{\rm c}^2$ from (6.2) with the replacement of $\alpha$ by $\beta$.

\begin{figure}
\begin{center}
\includegraphics[width=.9\textwidth]{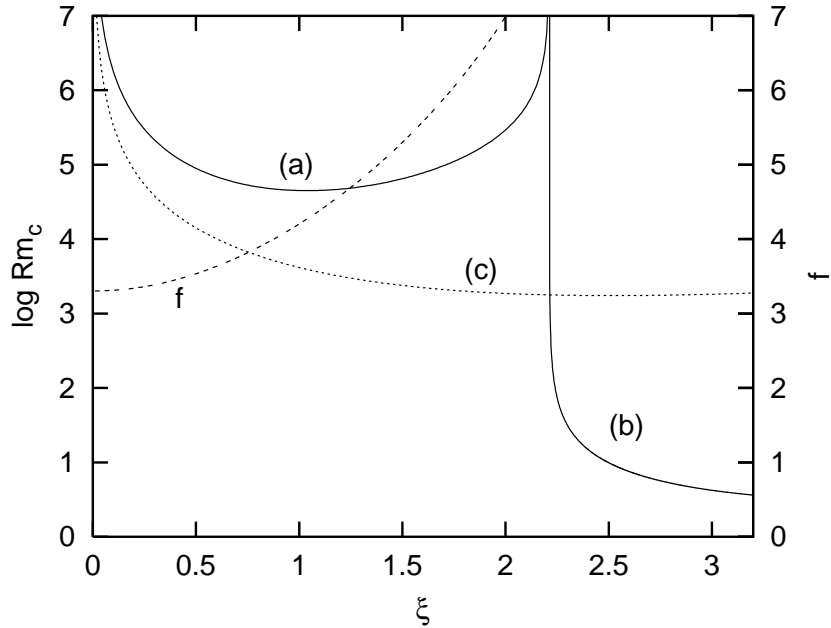}
\end{center}
\caption{Critical magnetic Rayleigh number and frequency of neutral oscillations versus magnetic
field strength. Curve $(a)$ -- oscillatory solution $Rm_{\rm c}^{\rm osc}$ and $(b)$ --
stationary solution $Rm_{\rm c}^{\rm st}$ with allowance for magnetophoresis and the negative
Soret effect ($\Psi=-0.6$); curve $(c)$ -- stationary solution $Rm_{\rm c}^{(0)}$ in the absence
of the particle diffusion.}
\label{f3}
\end{figure}

\begin{figure}
\begin{center}
\includegraphics[width=.9\textwidth]{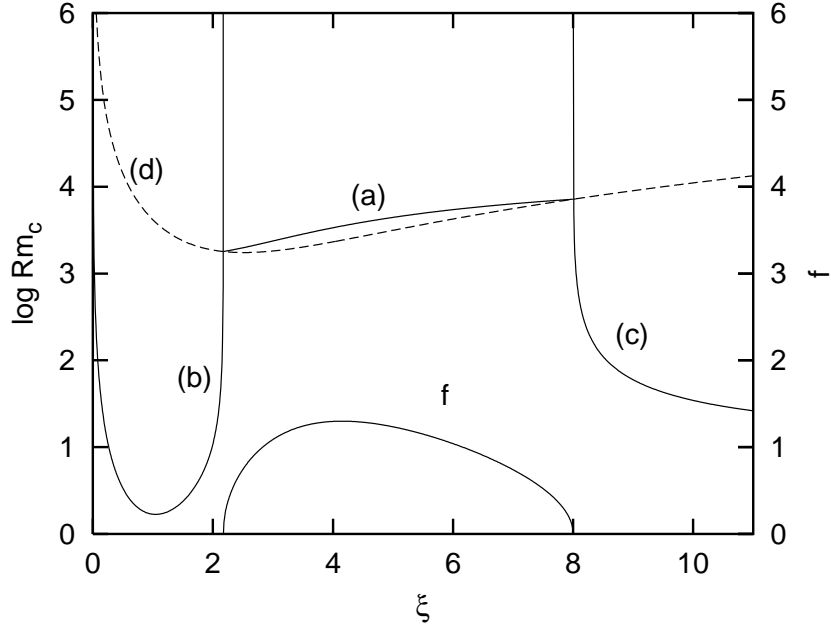}
\end{center}
\caption{Critical magnetic Rayleigh number and frequency of neutral oscillations versus magnetic
field strength. Curve $(a)$ -- oscillatory solution $Rm_{\rm c}^{\rm osc}$ and $(b),\,(c)$ --
stationary solution $Rm_{\rm c}^{\rm st}$ with allowance for magnetophoresis and the positive
Soret effect ($\Psi=0.6$); curve $(d)$ -- stationary solution $Rm_{\rm c}^{(0)}$ in the absence
of the particle diffusion.}
\label{f4}
\end{figure}

For a fixed negative separation ratio, $0>\Psi>-1$, oscillatory instability takes place within
an interval of the field $0<\xi<\xi_0$ (see Figs. 1 and 3), where $\xi_0$ satisfies the equation
$$\Psi +\xi{\cal L'}(\xi)/\xi=0\;.\eqno(6.3)$$
On the borders of the interval, the critical Rayleigh number (5.5) turns into infinity as
$\xi^{-2}$ for $\xi\rightarrow 0$ and as $(\xi -\xi_0)^{-1}$ for $\xi\rightarrow\xi_0$
(see Fig. 3). The frequency of neutral oscillations, $\omega_{\rm c}(\xi)$, commences from a
finite magnitude at $\xi=0$,
$$\omega_{\rm c}(0)=2\pi^2\sqrt{\frac{Pr\,|\Psi|}{1+Pr\,(1-|\Psi|)}}\;,$$
and grows monotonously with $\xi$ until it reaches $\xi_0$.

The upper border of the interval (5.3) is described by
$$\Psi(\xi)=\frac{\hat{\mu}-1}{\hat{\mu}}\,\xi{\cal L}(\xi)\;,\eqno(6.4)$$
where $\hat{\mu}=1+3(\mu_0-1){\cal L'}(\xi)$. The function (6.4) has a maximum $\Psi_m$ at a
certain value $\xi_m$ dependent on the initial magnetic permeability $\mu_0=1+4\pi\chi_0$. For
the fluid under consideration ($\mu_0=3$) we find $\xi_m=3.90\,,\;\;\Psi_m=0.806$. For any
positive $\Psi<\Psi_m$ there exists an interval $\xi_1<\xi<\xi_2$ of oscillatory instability --
see Figs. 1 and 4. The frequency of oscillations reaches a maximum within the interval and
turns into zero on its borders. Oscillatory branch $(a)$ of $Rm_{\rm c}(\xi)$ links in Fig. 4
two codimension-2 points, $Rm_{\rm c}(\xi_1)$ and $Rm_{\rm c} (\xi_2)$, in which stationary
branches $(b$ and $c)$ bifurcate.

A diagram of stability in the $(Rm_{\mathrm c},\;\Psi)$--plane at the fixed magnitude of the
field, $\xi=2$, is shown in Fig. 5. This diagram looks more or less traditional for binary
mixtures [14,15] with exception of two things. Firstly, the codimension-2 point (i.e., the
intersection point of $a$ and $c$ branches) is usually located in the close vicinity of
$\Psi =0$. In Fig. 5 this point has essentially shifted towards positive $\Psi$ {\em due to
magnetophoresis}. Secondly, the stationary branch $b$ as a general rule is located in the region
of negative Rayleigh numbers, i.e., the branch determines the onset of convection in the binary
mixture heated from {\em above}. Thermomagnetic mechanism of convection knows, however, neither
the top nor the bottom: the magnetic Rayleigh number $(3.11{\rm a})$ is proportional to
$(\Delta T)^2$, so it is always positive.

\begin{figure}
\begin{center}
\includegraphics[width=.9\textwidth]{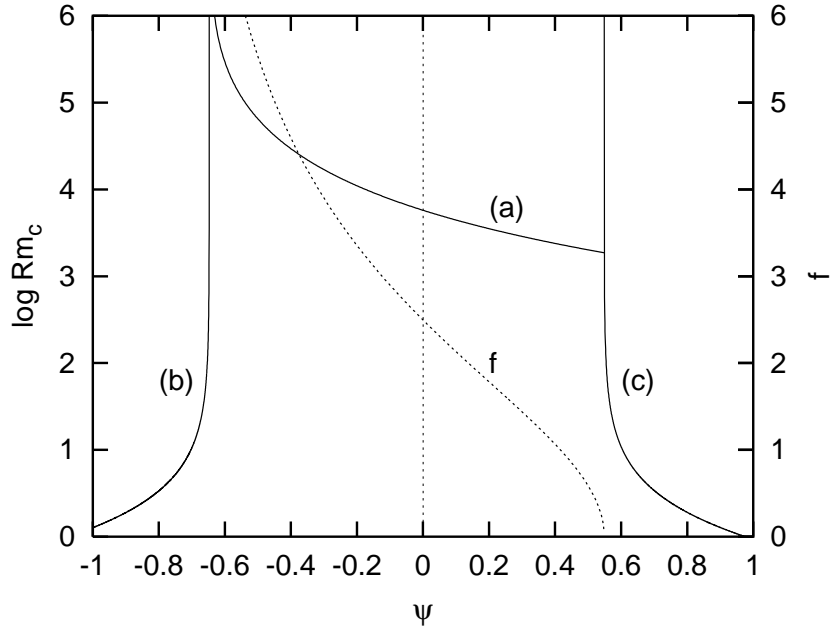}
\end{center}
\caption{Critical magnetic Rayleigh number and frequency of neutral oscillations versus
separation ratio at the fixed magnetic field strength $\xi=2$. Curve $(a)$ -- oscillatory
instability, curves $(b)$ and $(c)$ -- stationary instability.}
\label{f5}
\end{figure}

\begin{figure}
\begin{center}
\includegraphics[width=.9\textwidth]{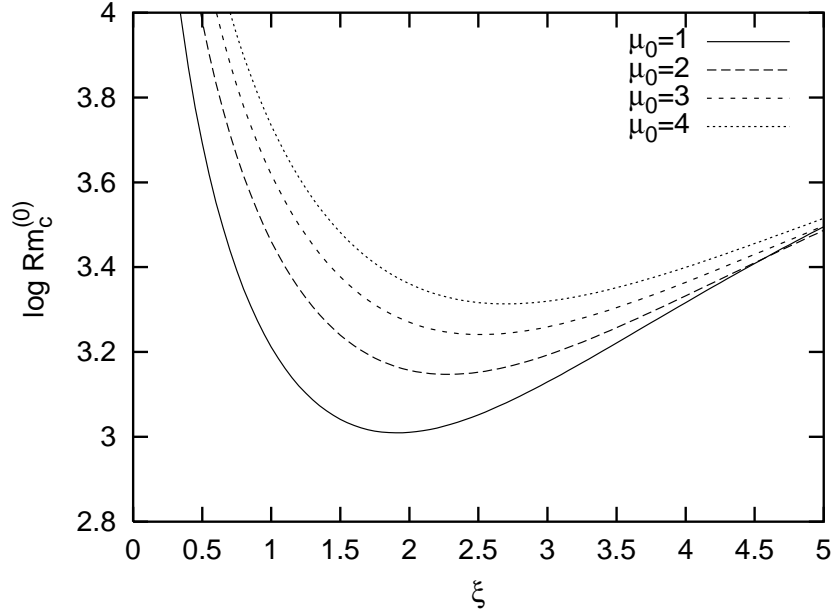}
\end{center}
\caption{Stationary instability curves for different $\mu_0$ in
the case when the particle diffusion is not operative.} \label{f6}
\end{figure}

Thus, convective instability of magnetized ferrofluids is strongly
effected by the magnetophoretic and thermophoretic (the Soret
effect) transfer of magnetic grains. The most interesting
consequence of the transfer is an opportunity to observe
oscillatory convective instability. However, as we have already
commented, to observe predicted oscillations one needs to satisfy
certain experimental conditions. Namely, the temperature
difference must not be increased faster than the limit imposed by
concentration diffusion. In the opposite case, a ferrofluid
behaves like a pure fluid what results in stationary instability
at the critical Rayleigh number $Rm_{\rm c}^{(0)}$; its dependence
on $\xi$ for some magnitudes of $\mu_0$ is shown in Fig. 6. In
conclusion, it is worth to remind that all our results are
relevant to thin ferrofluid layers $(\sim 1{\rm mm})$ where
magnetic mechanism of convection predominates over the buoyancy
mechanism, and the characteristic diffusion time, $\tau_{\rm D}$,
is not too long. To compare above-stated theoretical predictions
with future experimental results one should perform a more precise
analysis, which would satisfy realistic boundary conditions (4.1),
(4.2) and (4.8) instead of idealized boundary conditions (4.9)
used in the paper.
\subsection*{Acknowledgments}
I am grateful to the Alexander von Humboldt Foundation for a
Meitner-Humboldt research award and acknowledge the hospitality at
the Bayreuth and Saarlandes Universities. This work has been also
supported by a Grant 336/00--15.3 from the Israel Science
Foundation.

\section*{Glossary}
\begin{tabular}{l l}
$\phi$ & concentration of magnetic grains (volume fraction)\\
$\bf M$ & magnetization\\
$\bf H$ & magnetic field\\
$\bf B$ & magnetic induction: $\bf B=H+4\pi M$\\
$\bf F$ & magnetic force (volume density)\\
$\bf v$ & ferrofluid velocity\\
$T$ & temperature\\
$k_{\mathrm B}$ & Boltzmann's constant\\
$\rho$ & mass density\\
$\eta$ & fluid viscosity\\
$\chi_0$ & initial magnetic susceptibility\\
$\mu_0$ & initial magnetic permeability: $\mu_0=1+4\pi\chi_0$\\
$\chi$ & magnetic susceptibility: $\chi=M/H$\\
$\mu$ & magnetic permeability: $\mu=1+4\pi(M/H)$\\
$D$ & mass diffusion coefficient\\
$\kappa$ & thermal diffusivity\\
$S_{\mathrm T}$ & Soret coefficient\\
$\Psi$ & separation ratio for magnetic convection: $\Psi=(T/\phi)S_{\mathrm T}$\\
$t$ & time\\
$\tau_{\mathrm D}$ & characteristic diffusion time\\
$\tau_{\mathrm T}$ & characteristic thermodiffusion time\\
$Le$ & Lewis number: $Le=\tau_{\mathrm T}/\tau_{\mathrm D}$\\
$Pr$ & Prandtl number\\
$Ra$ & Rayleigh number\\
$Rm$ & magnetic Rayleigh number\\
$k$ & wave number\\
$\xi$ & Langevin's parameter (dimensionless magnetic field)\\
${\cal L}(\xi)$ & Langevin's function: ${\cal L}(\xi)=\coth\xi-\xi^{-1}$\\
$\lambda$ & increment of a normal mode\\
$\omega$ & angular frequency of neutral oscillations\\
$f$ & frequency of neutral oscillations: $f=\omega/2\pi$\\
\end{tabular}


\begin{thebibliography}{15.}
\addcontentsline{toc}{section}{References}

\bibitem{1}
M.I. Shliomis: Sov. Phys.-Uspekhi {\bf 17}, 153 (1974)
\bibitem{2}
R.E. Rosensweig: {\em Ferrohydrodynamics} (Cambridge University
Press, Cambridge 1985)
\bibitem{3}
V.M. Za\u{i}tsev, M.I. Shliomis: J. Appl. Mech. Techn. Phys. {\bf
9}, 24 (1968)
\bibitem{4}
B.A. Finlayson: J. Fluid Mech. {\bf 40}, 753 (1970)
\bibitem{5}
M.I. Shliomis: Fluid Dynamics {\bf 6}, 957 (1973)
\bibitem{6}
P.J. Stiles, M. Kagan: J. Magn. Magn. Mater. {\bf 85}, 196 (1990)
\bibitem{7}
C.L. Russell, P.J. Blennerhassett, P.J. Stiles: J. Magn. Magn.
Mater. {\bf 149}, 119 (1995)
\bibitem{8}
A. Recktenwald, M. L\"{u}cke: J. Magn. Magn. Mater. {\bf 188}, 326
(1998)
\bibitem{9}
J.-C. Bacri: private communication
\bibitem{10}
M.I. Shliomis: Sov. Phys.-JETP {\bf 34}, 1291 (1972)
\bibitem{11}
B.V. Derjaguin, S.S. Dukhin, A.A. Korotkova: Kolloidn. Zh. {\bf
23}, 53 (1961)
\bibitem{12}
L.D. Landau, E.M. Lifshitz: {\em Electrodynamics of Continuous
Media} (Pergamon Press, New York 1984)
\bibitem{13}
M.I. Shliomis, M. Souhar: Europhys. Lett. {\bf 49}, 55 (2000)
\bibitem{14}
D.T.J. Hurle, E. Jakeman: J. Fluid. Mech. {\bf 47}, 667 (1971)
\bibitem{15}
E. Knobloch, D.R. Moore: Phys. Rev. A {\bf 37}, 860 (1988)

\end{thebibliography}
\end{document}